# Requirements Volatility in Software Architecture Design: An Exploratory Case Study

Sanja Aaramaa
M3S
Faculty of Information Technology and Electrical Engineering
University of Oulu
Finland
sanja.aaramaa@oulu.fi

Sandun Dasanayake
M3S
Faculty of Information Technology and Electrical Engineering
University of Oulu
Finland
sandun.dasanayake@oulu.fi

Markku Oivo
M3S
Faculty of Information Technology and Electrical Engineering
University of Oulu
Finland
markku.oivo@oulu.fi

Jouni Markkula
M3S
Faculty of Information Technology and Electrical Engineering
University of Oulu
Finland
jouni.markkula@oulu.fi

Samuli Saukkonen
M3S
Faculty of Information Technology and Electrical Engineering
University of Oulu
Finland
samuli.saukkonen@oulu.fi

## ABSTRACT

Requirements volatility is a major issue in software (SW) development, causing problems such as project delays and cost overruns. Even though there is a considerable amount of research related to requirement volatility, the majority of it is inclined toward project management aspects. The relationship between SW architecture design and requirements volatility has not been researched widely, even though changing requirements may for example lead to higher defect density during testing. An exploratory case study was conducted to study how requirements volatility affects SW architecture design. Fifteen semi-structured, thematic interviews were conducted in the case company, which provides the selection of software products for business customers and consumers. The research revealed the factors, such as requirements uncertainty and dynamic business environment, causing requirements volatility in the case company. The study identified the challenges that requirements volatility posed to SW architecture design, including scheduling and architectural technical debt. In addition, this study discusses means of mitigating the factors that cause requirements volatility and addressing the challenges posed by requirements volatility. SW architects are strongly influenced by requirement volatility. Thus understanding the factors causing requirements volatility as well as means to mitigate the challenges has high industrial relevance.

## 1 INTRODUCTION

Requirements changes affect throughout all software (SW) development lifecycle, for example leading to higher defect density in testing phase [51]. Due to various internal and external factors, changes in individual requirements or in sets of requirements are inevitable. The changing nature of requirements in requirements engineering (RE) is denoted by requirements volatility. [10]. Since they are responsible for system structure, SW architects are the stakeholders that are greatly affected by requirements volatility [8], [17]. SW architecture design plays a prominent role in SW development, as it acts as the foundation of the SW systems and shapes the final outcome. SW architects must make critical decisions based on requirements [24]. Making sub-optimal architecture decisions can decrease system quality and cause problems later on [14]. Changes in requirements may necessitate the redesign of the SW to accommodate those changes, leading to an unstable SW system [26]. Correcting system shortcomings after completing system development is very complicated and far costlier than identifying and addressing them during initial design [51]. Existing literature discusses requirements volatility mainly from the project management viewpoint [40] and it has been studied related to SW maintenance [52] and coding [25]. But empirical evidence about requirements volatility from the SW architects' viewpoint and interplay between RE and SW architecture design in industrial environments is not extensively addressed.

To provide empirical insights into requirements volatility from SW architects' point of view, an industrial case study was conducted to explore requirements volatility in the context of SW architecture design. Fifteen SW architects involved in architecture design were interviewed for the case, the objective of which was to identify challenges that SW architects face due to requirements volatility and to propose means to address those challenges. The following research questions were derived from the objectives of the case.

RQ1: *What are the factors that cause requirement volatility?*

RQ2: *What challenges does requirements volatility pose to SW architecture design?*

RQ3: *What are means to address the identified challenges in SW development?*

This paper is structured as follows. Section 2 outlines related scientific literature. Section 3 describes the research approach. Section 4 provides an overview of the software development process in the case company and answers RQ1, describing the challenges identified based on the interviews. Section 5 discusses the means to address the identified challenges, based on existing scientific literature and the researchers' experience, thereby answering RQ2. Finally, Section 6 discusses conclusions.

## 2 RELATED WORK

RE is a systematic way to elicit, organise and document system requirements as well as a process to establish and maintain agreement between a customer and a systems provider [32]. Many authors emphasise the nature of change in the RE process [59], [35], [31]. In addition, the requirements management process is defined as a process of managing changes in the requirements [29]. Nurmuliani [43] defines requirements volatility as '*the tendency of requirements to change over time in response to the evolving needs of customers, stakeholders, the organisation and the work environment'*. Requirements volatility may also be defined as a change that could occur to a requirement [53]. This is in line with Nurmuliani's operationalised definition of requirements volatility [43]. Several internal and external factors, such as stakeholder feedback, technology, the market situation and customer needs, can cause changes in requirements [38]. Christel [10] groups requirements elicitation problems into three categories: scoping (information mismatch), understanding (inter- and intra-group) and volatility (requirement changes). The major reason for requirements volatility is changing user needs [48], which primarily lead to changes in individual requirements. Other important factors affecting volatility are conflicting stakeholder views and complexity of organisation [18], which lead to changes in the content of forthcoming releases. In addition, problems in understanding and scoping cause volatility [10]. Requirement information mismatch is also denoted requirement uncertainty [40], [39]. Existing literature on requirements volatility emphasises management viewpoint [58], [45]. These studies sought factors causing requirements volatility [10], its effects on projects [61] and means to mitigate those effects [1]. In addition, requirements volatility has been investigated in relation to SW maintenance [52] and coding [25], but not extensively in the context of SW architecture design. In short, requirements volatility is a phenomenon caused by various internal and external factors that lead to changes in the single requirements or sets of requirements. Changes in the sets of requirements have been claimed to results in more severe consequences than those in one requirement [16]. Applying an iterative RE process model has been proposed as a means to tackle requirements volatility [34]. For example, the twin peaks model is an iterative approach in which requirements and software architecture are developed in parallel from the very beginning of the software development life cycle [44].

*SW architecture* is the foundation that the SW system is built upon. The purpose of designing the SW architecture is to provide a unified vision about the system and improve understanding of its behaviour. The architecture, including diagrams, use cases and semantics, reduces ambiguities and shortens the time it takes for stakeholders to understand the constraints, behaviour, timing, layout for instance. Stakeholders involved in SW architecture design must make various decisions throughout the SW system life cycle regards to development, evolution and integration [24]. SW architecture design is inherently complex, and complexity is further increased because the architecture must address various stakeholder concerns [55] that may conflict with achieving SW system development goals. According to recent empirical studies on SW architecture design decision-making, in industrial environments, SW architects use experience, intuition and other informal approaches rather than using formal tools and techniques [15], [57]. SW architecture design is considered primarily as the SW architect's responsibility [13]; nevertheless, the active involvement of other stakeholders, such as SW developers, product managers and customers, is crucial for achieving the better understanding about the criteria the architecture must meet. An important element of the architecture design process is recording architecture design rationale, as understanding the reasons behind a certain architecture design decision can be critical during SW system maintenance and evolution [56]. Architectural technical debt refers to the consequences faced late in the SW development process due to sub-optimal architecture decisions and trade-offs [33], [14]. SW teams accumulate architectural technical debt due to their own actions and due to external events related to natural software aging and evolution. Even though technical debt related to coding issues can be detected using various tools, architecture technical debt mostly remains invisible and grows over time. [30] Factors that contribute to accumulating architectural debt include uncertainty about requirements at the beginning of development, the introduction of new requirements during the SW development process, time pressure, feature-oriented prioritisation and specification issues with critical architectural requirements [36].

The RE process and SW architecture design are called "twin peaks" and considered equally important [37]. Since they are closely connected, decisions made regarding one can affect the other [44]. Even though the traditional waterfall model leads to freezing requirements before moving into design and making hard to change architecture decisions, in reality the changes can occur in both areas and they can affect one another [11] Non-functional requirements (NFRs) constitute the majority of stakeholder concerns and greatly influence shaping up the architecture. The interactions between NFRs are among the main factors that should be taken into consideration during architecture design, as architecture either allows or precludes almost all NFRs of a system [12]. SW architecture design decisions are primarily based on Architecturally Significant Requirements (ASRs), which are critical to shaping system architecture [7]. While NFRs take the large share of ASRs due to their ability to affect the whole SW system, it can also include functional requirements. Correctly identifying and classifying architecturally significant functional requirements (ASFRs) is also critical for an architect to make informed decisions [41], [6]. The modern iterative software development approaches facilitate close interaction between requirements and architecture and help making rapid adjustments.

## 3 RESEARCH METHOD

The research method selected for this study was the case study, because the objective of the study was to research the requirements volatility in SW architecture design in the industrial setting [47]. The case study method is well suited for the researching real life phenomena in its natural setting [60]. Figure 1 depicts the phases of the case study.

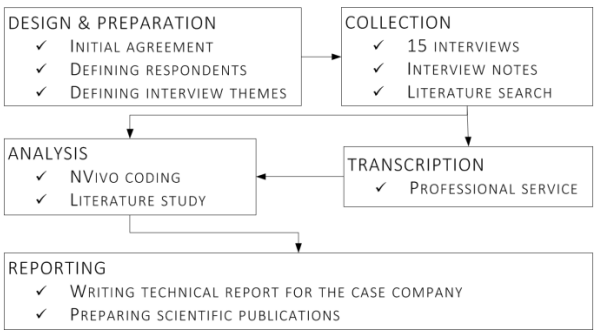

**Figure 1: The utilized case study process.**

*Case context*: The case study was conducted in a company with more than 900 employees in 25 offices worldwide. However, the product development is mainly done in three countries. The case company is a comprehensive SW solutions provider to both private and business customers and SW products for private customers. The SW solutions offered for business customers include a number of management tools and company services for the worldwide market. In addition, the case company provides various SW products for private customers to be used on an array of devices, ranging from desktops to handheld mobile devices. The SW development activities in the case company are carried out mainly in three countries.

In the case company there are three parallel business lines: U1, U2 and U3. The units are divided based on their business focus and each unit have their own financial responsibilities. There also is a horizontal business line, U4, which is responsible for providing services that are commonly shared by other three business units. The business lines operate as independent entities and within the business lines, there is a flat hierarchy of teams mostly organised based on projects. Along with the business units, a special unit, U5, consists of technical experts who take company-wide decisions related to technical matters. Individual teams are mostly self-organising and typically consist of four to eight people. While teams are free to operate according to their own agendas, they might have to interact and align with other teams, depending on the nature of the project. Some teams have their own architect or scrum master, but this is not the case for every team.

**Table 1. The details of the interviews.**

| ID | Unit | Responsibilities | The years of experience (in the case company) |
|---|---|---|---|
| I1 | U1 | SW dev./ Architect | 18 (17) |
| I2 | U1 | SW dev. | 7 (2) |
| I3 | U1 | SW dev./ Architect | 5 (3) |
| I4 | U1 | SW dev. | 20 (2) |
| I5 | U2 | SW dev./ Team lead | 15 (15) |
| I6 | U2 | SW dev./ Team lead | 20 (20) |
| I7 | U2 | SW dev./ Architect | 14 (8) |
| I8 | U3 | Program/ Project lead | 13 (7) |
| I9 | U3 | SW dev./ Architect | 13 (3) |
| I10 | U3 | SW dev. | 22 (15) |
| I11 | U3 | SW dev./ Architect | 13 (9) |
| I12 | U3 | SW dev./ Architect | 24 (4) |
| I13 | U4 | SW dev. | 15 (5) |
| I14 | U4 | SW dev. | 19 (6) |
| I15 | U5 | Lead architect | 15 (14) |

*Case study design and preparation.* An interview guide, which consisted of the set of open ended questions grouped into different themes, was prepared to guide the interviews. The interview guide consisted of the following themes; a general background, SW development process, requirements engineering and SW architecture design, SW architecture design challenges, solutions and expectations. The background questions covered the general information such as the experience of the interviewee, typical project and team composition as well as company description. The SW development process theme covered the overall development life cycle. The rest of the themes were designed to go into a deep discussion about requirement engineering and SW architecture design activities. The final part targeted, the various challenges faced by the interviewees and the possible solutions from their perspectives.

A pilot interview was conducted to test the interview guide as well as the interviewers approach. The participant of the pilot interview had a long history working in industry as a SW developer and a SW architect. In addition, the feedback received from the company representatives was also used to improve the interview guide. The reviewers of the questionnaire and the pilot interview participant did not participate in the actual case study.

*Data collection.* Fifteen semi-structured, thematic [20] interviews were conducted with SW experts during November and December 2014 as the primary method of data collection. All the interviewees acted as software architects in their respective teams even though not all of them are titled as architects (cf. Table 1). The interviewees were selected to represent all five units in the company's technical organisation. Interviewees have different levels of work experience, are active in various projects and are located at three company sites. The experience given in Table 1 reflects the years of experience in SW development in industry. The number without brackets is about experience in total while the one in brackets is for the years in the case company. Twelve interviews of 1–2 hours in length were conducted by two researchers face to face (F2F) at case company sites, and three interviews of the same duration were conducted via Skype. Interviewees were provided in advance with the list of interview questions, which was used as a guide, and more detailed questions emerged during the interviews.

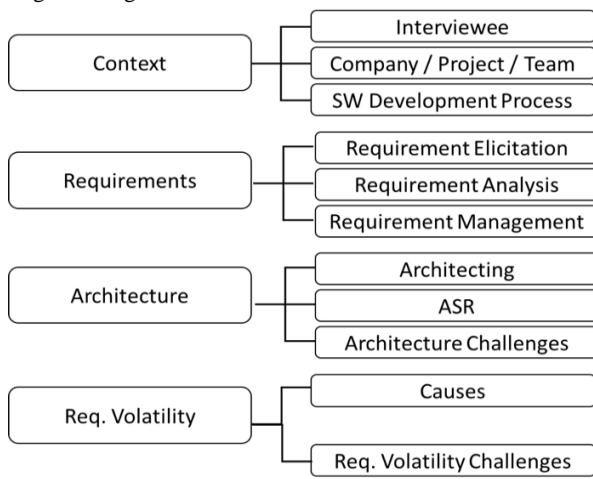

**Figure 2: The themes used in the data analysis.**

*Data analysis.* The interviews were professionally transcribed and fed into NVivo, a qualitative data analysis tool, as the starting point of the data analysis. The researchers went through each of the transcribed interviews and labelled the relevant information using a pre-defined set of themes based on the interview guide. Initially, there were the limited set of themes based on the research questions and as the process going forward, new themes were emerged and the sub themes were also created. The themes used during the data analysis are shown in Figure 2. The themes are not mutually exclusive and the same data can be labelled with multiple labels belongs to different themes.

This case study adapted the reporting guidelines provided by [47]. A strict privacy policy was followed that described all necessary elements of the case study while protecting the integrity of the company and individuals [4]. Interviewees' identities are protected by, for example, using aggregated information instead of presenting interview excerpts and by avoiding use of corresponding IDs in tables.

The results are based on the interviews and publicly available information about the company. The topics represent the aggregated viewpoints of the respondents throughout the interviews. The issues mentioned only by a single interviewee have not been included in the results, as they were considered not expressing shared understanding among SW architects. At first, an overview of SW development practices are outlined, which helps to understand the development context. This is followed by the description of requirements volatility challenges in the case company.

## 4 SW DEVELOPMENT PROCESS

SW development practices in the case company are quite informal and vary from team to team. Even though the company has a history of experimenting with various lean and agile SW development approaches on the company level, at the moment, there is no company-wide SW development approach. SW teams are free to select their own approaches unless there are specific restrictions such as customer preferences. SW teams tend to create their own approach by selecting and combining various agile and lean practices, including following sprints, maintaining product backlog and using Kanban boards. Although interviewees' have responsibility for SW architecture design, their involvement in RE is limited to occasionally providing expert opinion. Thus, interviewees' understanding about RE is not as extensive as is their knowledge of SW architecture design.

### 4.1 Requirements Engineering

SW architects are not directly involved in customer requirement elicitation and analysis. In the case company, *elicitation* is accomplished using techniques such as focus groups, beta testers, and direct customer communication. Occasionally, customers' ideas are expressed at so abstract a level that they can hardly be translated as requirements. On the other hand, it is possible that requirements will be stated as full-scale technical specifications. Customer needs are clarified during the requirement *analysis*. The SW product owner (PO) is a link between the SW team and product management (PM). The PO has the responsibility to communicate with the PM to clarify what must be implemented. For the majority of interviewees, the PO is the sole connection point to requirements elicitation and analysis phases. During the clarification process, SW architects are occasionally consulted to define the technical feasibility of requirements and choose the best implementation solution. At this phase, requirements volatility factors include changing customer needs and evolving technological understanding.

A project management tool (PMT) is the main medium for *documenting* requirements, which usually are expressed as features or backlog items. In addition to the primary PMT, legacy and team-specific tools and sticky notes are used to communicate tasks, store customer information and document decision rationales. The level of details in information on requirements varies depending on a product and a team. Sometimes, only a feature name is recorded, but at the other extreme, descriptions include even the contact information of the relevant technical specialist on the customer side. In the case of private customer products, requirements are created by experts based on a foreseen market. Usually, the creator of a requirement is recorded, but sometimes it is not known whether the requirement originated with a customer or an internally identified technology gap. Interviewees related that they sometimes needed more technical details or contextual descriptions to be able to choose the best implementation alternative. Big corporate customers may have strict requirements about formal documentation to be delivered to the customer. Interviewees pointed out that a requirement description is always a compromise between level of detail and time available for the task.

The PM is responsible for *prioritising* requirements, taking into account, for example, company strategies and the importance of the customer. The dominant factor when setting requirement priority is the customer: The bigger the customer, the higher the priority of its requirements. Even though some features are technically feasible and could contribute improving the product quality in the long run, it may be very difficult to say no to features that a customer wants, especially if the customer is big. Other factors taken into account when setting requirement priorities include development cost, feature size, product roadmap, criticality and external audit results, which are publicly available and used to rank the solution providers in the domain. Interviewees were involved in requirement prioritisation, proposing product improvements and project scoping meetings. Most interviewees mentioned having faced challenges with changing requirement priorities. As the backlog is updated frequently, changing priorities contribute to requirements volatility.

## 4.2 Software Architecture Design

Since the majority of the teams follow agile and lean approaches, the design and implementation are done iteratively, leading to a shorter design phase than in the traditional waterfall approach.

The SW architecture design process typically starts with backlog review meetings between the team and other relevant stakeholders. The objective of these meetings is to reach a consensus what needs to be developed to fulfil requirements. While the team is generally represented by senior members during the initial meeting, it is possible that the whole team is involved from the beginning. Once the basic ASRs are understood, the team creates a design proposal, which is delivered for review. The review is done at different levels, depending on the scale of the project or its dependencies to other products. Once decisions are made, the team is free to begin development and has the flexibility to make minor changes to the design. If the design must be altered considerably, the evaluation of the alterations is escalated. SW architecture designs and decisions taken during discussions at various levels are recorded using several methods. Even though the interviewees claimed that they have maintained some type of design documentation, attention to documenting design appears to be inadequate. The majority of teams use tools such as Wiki or PMT instead of traditional design documents to store their architecture decisions.

## 5 REQUIREMENT VOLATILITY – ORIGINS AND CHALLENGES

Through data analysis, it was possible to identify factors that cause requirements volatility in the case company as well as the challenges that requirements volatility poses to SW architecture design. This section first describes the factors of requirements volatility identified in the case company and then the consequences of requirements volatility. Thus the subsection **Error! Reference source not found.** answers RQ1 and the subsection **Error! Reference source not found.** answers RQ2. Quotes from interviews provide insights for collected empirical evidence. Each quote is from a different interview. However, to protect the integrity of the respondents the quotes are not labelled by the respondent ID. None of the respondents was a native English speaker. Therefore, it has been necessary to correct some minor language errors to ensure a proper message.

### 5.1 Factors Causing Requirements Volatility

Several factors that contribute to requirements volatility were identified in the case company.

*5.1.1 Requirement Uncertainty*. An important factor for causing requirements volatility is the uncertainty of requirements, which is realised for example as inadequate descriptions of backlog items. Often, features or backlog items lack detailed information; for example, a backlog item may have only a name, but no one knows why the item is there. The tool includes a customer acceptance criteria field as part of the requirement description. Most of the time, something is recorded in this field. However, often, the description is a couple of lines of text at an abstract level. This means that architects and testers must guess what must be fulfilled.

*"It* [description] *can be just couple lines of text and that's all and we need to guess what shall we do…. quality engineers always complain about it because they don't know how to test because it's not so clear how it should work."*

*5.1.2 Changing User Needs*. The case company provides multiple SW products for various customer groups and frequently comes across changing customer needs. As most of the requirements for private customer products are decided within the company, corporate customers are the main source of requirements changes. The long-term business relationship between corporate customers and the company makes it difficult to refuse to adapt to changing customer needs.

*"Well, since we are doing this project with, constantly changing requirements. I don't see much chance for improving the process because we are just, basically adapting. And not planning ahead."*

*5.1.3 Dynamic Business Environment*. The company operates in a dynamic business domain and must adjust its strategies for accommodating development in that domain to stay ahead of the competition. The severe market situation requires constant changes in requirement priorities. As most of the company's private customer products runs on smartphones where the operating systems are highly fragmented and subjected to frequent changes, the company has to make frequent changes to their products to accommodate those changes.

*"This list we see for quarter is something that we can work on. Whatever in future is, at least, that is subject to change because market change, situation change and stuff like that. So we wouldn't know."*

*5.1.4 Stakeholder Dependencies.* The company is structured along business lines, each of which runs its projects independently. However, sometimes delivering a solution requires collaboration among teams from different business units. For example, a team that has developed a mobile application might have to interact with teams that have developed the same application for different platforms and with teams that provide server-side support for those applications. In this situation, changes in requirements in one unit lead to changes in another one.

*"When we have external dependencies on the teams in, especially if they are another location, it's sometimes quite hard to make sure that everything happens in time."*

*5.1.5 Communication Issues.* As the case company has a globally distributed customer base, multiple development sites and virtual SW teams, it is challenging to communicate requirements. Communication issues may begin during the elicitation and customer negotiation phases. Typically, this is due to the fact that the terminology and semantics differ between customers and developers. These differing domains bring to the table different terminology and concepts. Later in the process, SW teams may face language barriers and cultural differences that pose communication challenges. Communication issues are present, even though the company has tried several supporting tools and approaches to improve communication both within the company and with customers.

*"And we often need clarification all the way from customers…, but we do have this kind of feedback cycle from us to the customers, so that we can find out what exactly is wanted. Because that's really not that clear always."*

## 5.2 Requirements Volatility Challenges in Software Architecture

Interviews revealed several challenges that requirements volatility poses to SW architecture design. It should be noted that requirements volatility is not the sole reason for these challenges, but requirements volatility is the scope of this study.

*5.2.1 Scheduling.* Typically, requirement clarification takes so much time in the case company that architects receive requirements very late in the project. This causes challenges in scheduling both on the team level and the organisational level. On the team level, the later the architects receive the requirements, the less time they have to design architecture, and some steps must be omitted. The company creates development plans quarterly. The aim is to have two forthcoming quarters planned to provide an overall idea about what should be achieved in six months. However, quarterly plans are subject to change. Often, when a quarter begins, the schedule applies for only a couple of weeks, and then the content must be re-planned. The most important reason for scheduling issues is requirements volatility. In the worst cases, priorities change daily, in which case, architects have no choice but to work on the item at hand and then take the next one on the backlog list, which might be different the next morning.

*"Actually, we can't use Scrum basically because our priorities change all the time, like.. maybe, sometimes daily so, yeah, we basically get the next item and work on it."*

*5.2.2 Synchronisation.* Stakeholder dependencies are one factor causing requirements volatility, which, in turn, causes synchronisation issues. Interviewees noted that sometimes they are unable to deliver products on time due to delays in other units. Beside inter-unit synchronisation issues, teams in the same business unit but at different development sites have synchronisation challenges caused by lack of physical proximity. Synchronisation needs and development dependencies also influence the tools used. PMT was used for project management, requirement descriptions and bug fixing. These activities require very different tool functionalities, which are supported only partially. For example, maintaining the backlog within a project and following feature development work well, but cross-project management, such as moving a feature from one project to another, is not supported.

*"Probably the most complex thing is that, we need to somehow synchronize the requirements between the different teams and, that's why, having some leadership team would be beneficial because they would synch up together what they are gonna do, what resources they have, how they would transfer their stories between the teams and et cetera."*

*5.2.3 Architectural Technical Debt.* The company's requirement prioritisation criteria are strongly business driven, favouring market needs over architectural considerations. Overlooking architectural aspects when prioritising requirements accumulates architectural technical debt. As architects are overwhelmed with volatile requirements, they are not necessarily able to find out the optimal architectural design choices. Specifically, prioritising functional requirements over NFRs is a major issue, as NFRs are the majority of ASRs, and neglecting them leads to sub-optimal architecture design.

*"I think the biggest driver usually for getting something prioritized really fast is money. So if the customer is big enough and the expected income is big enough,* [case company] *will run through hoops…, for smaller customers even if the smaller customer is asking for something that makes much more sense. So I think that, the primary driver is economic. So instead of doing feature development, we kind of get overridden by the* [business customer] *deliveries all the time, because that's where the money comes from."*

*5.2.4 Tracing Design Rationale.* The interviewees understand the importance of recording design rationale while making an architecture decision and tracing back to it during the later stage of the software development process. However, they find it difficult to maintain design documents, the most common way of recording design decisions, because they require frequent updates due to the volatile requirements. Even though PMT supports current need to some extent, there are major issues with outdated and unstructured information, broken links and inability to trace decision rationale. Typically, PMT is not used to store customer-related information, since it is considered too insecure. Thus, there must be some other means to record customer data.

*"So usually the reasoning* [behind design decisions]*, happens, it's kind of like corridor discussion where we have a meeting where we talk about it and then we report what we decided to do in the PMT items but of course there's lot of stuff that we miss, so we don't really document the why, we document the what. And of course these discussions in the meeting, kind of contain the why also but if you are not in the meeting then that information is not available, usually."*

The following Figure 3 summarises the identified factors causing requirements volatility in the case company as well as the challenges it poses to SW architects during the definition of SW architecture.

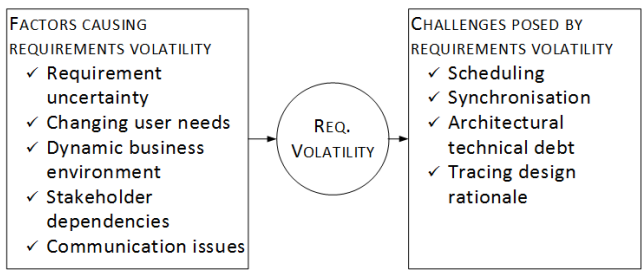

**Figure 3: The factors and challenges causing requirements volatility.**

## 6 ADDRESSING REQUIREMENT VOLATILITY CHALLENGES

This section discusses the requirements volatility challenges in SW architecture design in relation to existing scientific knowledge and in the light of the empirical data gathered from the case company, thereby answering RQ3. In general, there are two main approaches to addressing requirements volatility challenges: mitigate their causes or find means to manage their consequences to SW architecture design.

*Mitigating the causes of requirements volatility.* Interviewees provided contradictory opinions about how much information is available to them. On one hand, it was reported that at the beginning of a project, a significant amount of time is spent negotiating technical feasibility and clarifying actual needs, the intended behaviour of the product and dependencies with other features. On the other hand, it was noted that if the available information is not too detailed it leaves room for creativity and allows discovering the best technical solution. Missing requirements early in the process causes the costliest fixes later in the development process [43]. The starting point for addressing *requirement uncertainty* is to evaluate what information is crucial to whom, why and when. This should be stated explicitly in the information fields provided for describing the requirement. Unnecessary default requirement fields should be removed. However, this is not sufficient, since the quality of the descriptions depends on the expertise of the writer. According to interviewees, POs or marketing personnel do not have enough technical understanding about the product to write detailed requirement descriptions. In addition, it would be a waste of time to write an extensive requirement description just to find out later that the requirement is not technically feasible. The twin peaks approach where requirements and architecture are developed in parallel could be used to address this issue [44]. The impact caused by the lack of technical understanding of the people responsible for eliciting requirements can be mitigated through supportive means. One such way is asking probe questions, to identify ASRs from software requirement specifications [5], [6].

Maintaining a strong, mutually beneficial relationship between the customer and the development team is crucial to successfully managing *changing customer needs* [22]. While this helps to understand the customers' true needs and derive well-defined requirements from the beginning, it also allows developers to communicate the consequences of accommodating changes rather than blindly accepting them. Approaches to mitigate the effects of changing customer needs include eliciting gaps in requirement changes [46] and reusing existing requirements to identify the gap between elicited requirements and true user needs [2].

Following shorter development cycles can help address requirement prioritisation issues caused by a *dynamic business environment* [23]. However, following short design cycles can affect project scheduling and synchronisation. So it should be carefully considered.

As the SW teams in the case company sometimes depend on each other's work, their requirements are interconnected. Therefore, it was alarming to observe that identifying the *dependencies between the requirements of various stakeholders* sometimes was more wishful thinking than practice. Managing requirement dependencies is acknowledged as a challenging task [9] that may be addressed by supporting impact analysis [58]. One approach could be to establish a product management team across business units that would have overall responsibility for managing projects that span business units, including resources, scheduling, priorities and such.

Introducing advanced collaborative and communication mechanisms can overcome some of the *communication issues* among distributed software teams [3] as well as customer communication issues [28]. At the same time, using multiple communication media rather than a single communication channel can help avoid misunderstandings caused by cultural and language differences [50]. In the context of the case company, the personnel responsible for customer communication play a crucial role in effective communication.

*Managing the consequences of requirements volatility.* As previous sections have described, there are no direct causal relationships between one factor and one consequence, but relationships can be recognised. Therefore, there is no definite solution to address each identified consequence. As *scheduling* issues appear at the team level and at the organisational level, addressing them must be undertaken at both levels. At the team level, the situation could be improved by, for example, assessing the suitability of the elicitation techniques used and the adequacy of the requirement information collected. According to Hickey and Davis, an elicitation technique should be chosen based on problem, solution and project domain characteristics as well as known requirements [21]. On the organisational level, one solution could be to include a sufficient buffer [19] for planned releases as a response to requirements volatility. According to feedback from interviewees, interaction among team members located at various sites is not adequate, despite using various communication tools. When it comes to distributed teams, just maintaining work communication among team members is insufficient. The performance of distributed teams is affected by networking within the team and trust among members [54].

Lack of visibility among business units was mentioned constantly during interviews as hindering *synchronisation* among business units. While separations among business units may be necessary to organisational management, they cause several negative results in SW architecture design, the main one being the possibility of duplication of work, as the teams are not aware of each other's work. Considering the amount of human resources and talent in the case company, there are good opportunities for knowledge-sharing among engineers. Even though the technology council and company-wide steering group can prevent the large-scale duplication of effort, work still can be duplicated on the micro-level. Closer interaction among architects in various business lines will facilitate the identification of resources suitable for a given task and, hence, get it done more efficiently. Since individual business lines evaluate their own performance, collaboration with other business lines might not be high on their agendas. However, in the long run, business lines and the company as whole can benefit from a transparent approach.

Taking the views of software architects and developers into consideration during the prioritisation process can contribute to reducing *architectural technical debt* considerably. Since it is the SW architects' responsibility to recognise the ASRs that include NFRs and their effects on overall SW system architecture, architects are in the best position to identify factors causing debt and manage them to minimize accumulation of architectural technical debt [57]. In the context of iterative SW development, addressing architectural technical debt as a separate backlog item can ensure that it is addressed properly, as otherwise, the cross-cutting nature of NFRs makes it difficult to address them properly at any given point [42]. Improving traceability across the SW development process is important to understand the implications of changing requirements for SW architecture and assessing possible architectural debt [27].

SW architects should be encouraged to record design rationale by being given adequate tools [56]. In addition, a company-wide methodological approach to recording design rationale and maintaining necessary documentation should improve the quality of documentation and traceability of design rationale. Using the set of tools, including PMT, Wiki and a version control system, is a good move, as it helps maintain consistency and makes tool maintenance and support easier. However, to get the maximum benefit from the tools, it is important to educate engineers and provide necessary training. In addition, filling gaps in the existing tool chain or introducing a new tool chain that provides end-to-end tool solutions would help address identified challenges.

## 7 THREATS TO VALIDITY

According to Yin a construct and external validity as well as reliability are necessary conditions that have to be taken into account when conducting case studies. Internal validity has to be considered when conducting exploratory case studies. [60] Yin suggests using multiple sources of evidence, establishing the chain of events and having key informants to review case study report as tactics for ensuring construct validity in case studies. Internal validity can be addressed for example through considering a rival explanation and using logical models.

In this case study, threats to construct validity were mitigated by interview guide reviews done by professors and representatives of the case company. At the beginning of each interview, key terms related to the study were defined and discussed to ensure a common vocabulary among researchers and interviewees. In addition, a pilot interview was conducted to get feedback from an expert. The case study results were presented in the case company in a workshop, where practitioners had an opportunity give feedback to the researchers about the study results. The case study report was delivered to the case company representatives, including interviewees, and they were asked point out any corrections needed for the report.

Threats to internal validity must be taken into account when studying causal relationships. This study aimed to explore the challenges to SW architecture posed by requirements volatility. Since SW development is affected by several other factors, too, there is no clear causal relation between requirements volatility and SW architecture challenges. The examples of these factors are technological changes and company strategies. However, this study addressed requirements volatility only.

External validity relates to the generalisability of results. Traditionally, it has been suggested that the generalisability of results from a single case study is rather poor. [47] However, the results of case studies may be extended to other cases that have common characteristics. [47] Seddon & Scheepers, suggest that generalisation of results can be done based on a single case study as long as 1) a sample is carefully analysed, 2) relevant factors, which are true in a sample can be argued being true in larger similar context and 3) researchers seeking to generalise results discuss their findings in relation to prior studies. [49] Considering the context of the study it is expected that similar finding can be drawn when the following characteristics are present: a) globally distributed software development teams, b) a company operating in a dynamic market, c) a large company structured as autonomous units and d) serving a diverse customer base. Threats to external validity were taken into account by collecting data that can be used to characterise the subjects and case context. Examples of these data are experience of the interviewees in their field and in the company, team sizes, organisational structures and roles and the responsibilities related to them.

Threats to reliability relate mostly to means of collecting and processing data. These threats were addressed by the review of the interview guide and by agreeing upon a coding scheme prior to analysis.

## 8 CONCLUSIONS AND FUTURE WORK

This industrial case study was conducted to explore the challenges that requirements volatility poses to SW architecture design. Fifteen SW experts involved in SW architecture design in various business units were interviewed using a semi-structured interview as a guide.

This study revealed factors causing requirements volatility as the well as the challenges posed by the requirement volatility in the case company, which provides SW solutions for companies and SW products for consumers in a global market. Through the study important challenges that requirements volatility poses to SW architecture design were identified. Finally, the means to address the identified challenges were discussed. Some factors, such as requirements uncertainty or missing information, the constant change of priorities and shifting and competing goals, are recognised as preventing architects from analysing available options and taking the optimal course of action in complex, real-world scenarios. As discussed, the context in which architecture is designed is very demanding, and architects prefer to compromise quality rather than increase overhead. However, there is a possibility that the same factors that are considered advantages also affect architecture design negatively. For example, even though light-weight documentation is considered an advantage, it can contribute to requirements volatility and increase the risk of creating architectural technical debt.

As software engineering researchers are increasingly interested in the "twin peaks" of the software development: requirements and architecture design. This study provided empirical evidence about the relationship between them and how the process changes in one can affect another. The ultimate goal was to understand the complexity of the development environment and issues the practitioners face daily and thus propose feasible solutions for industry. This case study provides an example for practitioners how research may help to expose challenges, their reasons and impacts in the company. Practitioners may consult the results of the case study to identify similarities and differences in their practices. This in turn helps to find improvement directions.

In future research, another case study will be conducted in a company of different size and in a different domain to investigate whether the same challenges are present there. Cross analysis between the case studies will provide new insights and help increasing the generalisability of the findings. Based on the results of those case studies, it is planned to develop a framework that provides means for practitioners to identify the presence of challenges posed by requirement volatility and take necessary steps to mitigate the risks.

## ACKNOWLEDGMENTS
This work is funded by ITEA2 and Tekes - The Finnish Funding Agency for Innovation, via the MERgE project.